\documentclass[aps,prl,superscriptaddress,twocolumn]{revtex4-2}

\usepackage[utf8]{inputenc}
\usepackage{amsmath} 
\usepackage{amssymb} 
\usepackage{amsfonts} 
\usepackage{graphicx}   
\usepackage{verbatim}  
\usepackage{color}     
\usepackage{hyperref}   
\usepackage{natbib}
\usepackage{gensymb}
\usepackage{color}
\usepackage{multirow}
\usepackage{xfrac}

\begin{document}

\title{Carrier-Density Control of the Quantum-Confined 1$T$-TiSe$_2$ Charge-Density-Wave}

\author{T. Jaouen}
\altaffiliation{Corresponding author.\\ thomas.jaouen@univ-rennes1.fr}
\affiliation{Univ Rennes, CNRS, (IPR Institut de Physique de Rennes) - UMR 6251,  F-35000 Rennes, France}
\author{A. Pulkkinen}
\affiliation{D{\'e}partement de Physique and Fribourg Center for Nanomaterials, Universit{\'e} de Fribourg, CH-1700 Fribourg, Switzerland}
\affiliation{New Technologies Research Centre, University of West Bohemia, CZ-30100 Pilsen, Czech Republic}
\author{M. Rumo}
\affiliation{D{\'e}partement de Physique and Fribourg Center for Nanomaterials, Universit{\'e} de Fribourg, CH-1700 Fribourg, Switzerland}
\affiliation{Haute école d'ingénierie et d'architecture de Fribourg, CH-1700 Fribourg, Switzerland}
\author{G. Kremer}
\affiliation{D{\'e}partement de Physique and Fribourg Center for Nanomaterials, Universit{\'e} de Fribourg, CH-1700 Fribourg, Switzerland}
\affiliation{Institut Jean Lamour, UMR 7198, CNRS-Université de Lorraine, Campus ARTEM, 2 allée André Guinier, BP 50840, 54011 Nancy, France}
\author{B. Salzmann}
\affiliation{D{\'e}partement de Physique and Fribourg Center for Nanomaterials, Universit{\'e} de Fribourg, CH-1700 Fribourg, Switzerland}
\author{C. W. Nicholson}
\affiliation{D{\'e}partement de Physique and Fribourg Center for Nanomaterials, Universit{\'e} de Fribourg, CH-1700 Fribourg, Switzerland}
\affiliation{Fritz-Haber-Institute der Max-Planck-Gesellschaft, Faradayweg 4-6, 14195 Berlin, Germany}
\author{M.-L. Mottas}
\affiliation{D{\'e}partement de Physique and Fribourg Center for Nanomaterials, Universit{\'e} de Fribourg, CH-1700 Fribourg, Switzerland}
\author{E. Giannini}
\affiliation{Department of Quantum Matter Physics, University of Geneva, 24 Quai Ernest-Ansermet, 1211 Geneva 4, Switzerland}
\author{S. Tricot}
\affiliation{Univ Rennes, CNRS, (IPR Institut de Physique de Rennes) - UMR 6251,  F-35000 Rennes, France}
\author{P. Schieffer}
\affiliation{Univ Rennes, CNRS, (IPR Institut de Physique de Rennes) - UMR 6251,  F-35000 Rennes, France}
\author{B. Hildebrand}
\affiliation{D{\'e}partement de Physique and Fribourg Center for Nanomaterials, Universit{\'e} de Fribourg, CH-1700 Fribourg, Switzerland}
\author{C. Monney}
\affiliation{D{\'e}partement de Physique and Fribourg Center for Nanomaterials, Universit{\'e} de Fribourg, CH-1700 Fribourg, Switzerland}

\begin{abstract}
Using angle-resolved photoemission spectroscopy, combined with first principle and coupled self-consistent Poisson-Schr{\"o}dinger calculations, we demonstrate that potassium (K) atoms adsorbed on the low-temperature phase of 1$T$-TiSe$_2$ induce the creation of a two-dimensional electron gas (2DEG) and quantum confinement of its charge-density-wave (CDW) at the surface. By further changing the K coverage, we tune the carrier-density within the 2DEG that allows us to nullify, at the surface, the electronic energy gain due to exciton condensation in the CDW phase while preserving a long-range structural order. Our study constitutes a prime example of a controlled exciton-related many-body quantum state in reduced dimensionality by alkali-metal dosing.
\end{abstract}

\date{\today}
\maketitle

Layered metallic transition-metal dichalcogenides (TMDs) form the ideal platform for studying charge density wave (CDW) instabilities at the quasi-two dimensional (2D) limit as well as their interplay with Mott states and superconductivity \cite{Morosan2006a, Sipos2008, Kusmartseva2009a}. The 2 $\times$ 2 $\times$ 2 commensurate CDW that occurs at $\sim$200 K in the prominent TMD 1$T$-TiSe$_2$ is a representative case of a many-body state whose strong entanglement of the electronic and structural parts of the order parameter makes the determination of its driving force highly complex. For decades, the origin of the CDW has been under debate between a purely electronic excitonic insulator (EI) scenario and the Peierls mechanism of electron-phonon coupling \cite{Rossnagel2011}. More recently, a cooperative exciton-phonon mechanism has emerged as the most likely explanation of the CDW phenomenology in 1$T$-TiSe$_2$ \cite{vanWezel2010, vanWezel2010c}, encompassing both observed phonon and exciton softenings \cite{Weber2011b, Kogar2017b}. Yet, despite the increasing number of theoretical approaches employing such a scenario \cite{vanWezel2010,vanWezel2010c, Zenker2013, Watanabe2015, Kaneko2015, Kaneko2018}, experimental attempts of disentangling the coupled electronic and structural facets of the CDW state mainly focus on the ultrafast time domain and their distinct out-of-equilibrium dynamics \cite{Huber2022, Cheng2022, Zhang2022, Otto2021, Burian2021, Duan2021, Mulani2021, Lian2020, Hedayat2019, Monney2016, Porer2014}.

In this letter, we use a thermodynamic phase control approach, namely electron doping brought on by low-temperature alkali atoms adsorption at the 1$T$-TiSe$_2$ surface. Using angle-resolved photoemission spectroscopy (ARPES), first principle and coupled self-consistent Poisson-Schr{\"o}dinger (PS) calculations, we demonstrate that the electron-doped 1$T$-TiSe$_2$ surface supports a two-dimensional electron gas (2DEG) coexisting with the quantum-confined CDW. The carrier density within the 2DEG further serves as a tuning knob to control the Coulomb screening in the surface quantum well and as a means to follow a doping-driven thermal separation of the excitonic and structural subsystems of the 2D CDW \cite{Li2015, Rossnagel2010}. Our study highlights how the electronic band structures of exciton-related materials are affected by electrical tuning induced by alkali-metal dosing.

\begin{figure*}[t]
\includegraphics[scale=1]{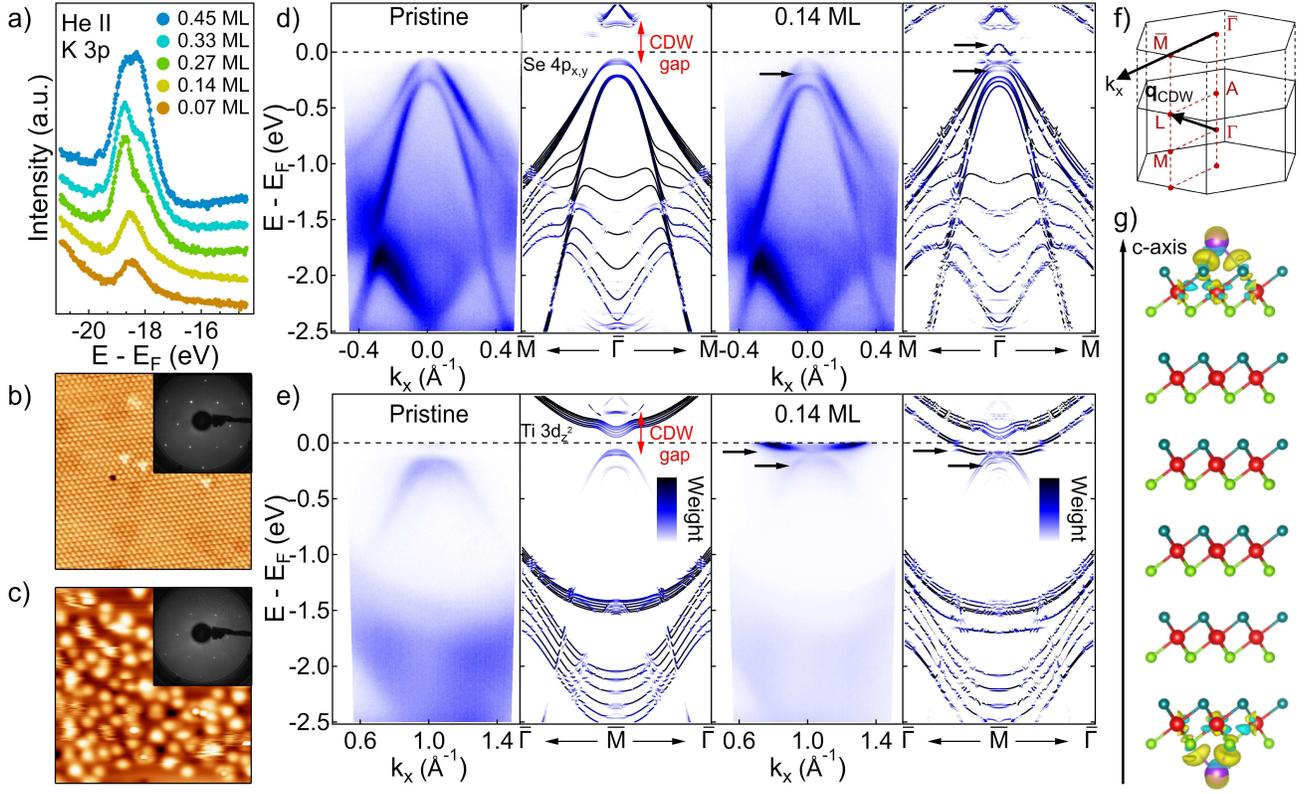}
\caption{a) K 3$p$ core-level evolution upon sequential K depositions. $h\nu$=$40.8$ eV. b)-c) 12$\times$12 nm$^2$ constant-current STM images of the pristine b) and of a 0.33 ML K-covered c) surfaces. V$_{\text{bias}}=-1$ V, I $=0.2$ nA, T = $4.6$ K. Also shown, as insets, are the corresponding LEED patterns ($E_p =$ 121 eV). d)-e) Comparison of large-energy scale ARPES intensity maps with band structures obtained from DFT slab calculations along the $\bar{M}$-$\bar{\mathrm{\Gamma}}$-$\bar{M}$ and $\bar{\mathrm{\Gamma}}$-$\bar{M}$-$\bar{\mathrm{\Gamma}}$ directions of the surface Brillouin zone (BZ) for both pristine (left-hand side) and K-covered (right-hand side) surfaces in the CDW phase. The horizontal black arrows indicate the QWS (see text). The color scales from white to dark blue indicate the spectral weights obtained using the unfolding procedure (see SM). $h\nu$=$21.2$ eV, T = 40 K. f) 3D BZ of 1$T$-TiSe$_2$. Also shown is one CDW $q$-vector, $\textbf{q}_{CDW}$. The $k_x$ axis depicted by the black arrow is the $\bar{\mathrm{\Gamma}}$-$\bar{M}$ direction of the surface BZ. g) Side view of the 1$T$-TiSe$_2$ 6-layers slab with a 2 $\times$ 4 K adlayer on the surface. The light blue isosurface refers to the missing charge and the yellow isosurface to the gained charge for a 0.0014 $\sfrac{e^-}{a_0^3}$ level, where $a_0$ is the Bohr radius.}\label{fig1}
\end{figure*} 

The 1$T$-TiSe$_2$ single crystals were grown by chemical vapor transport at 590 \celsius , therefore containing less than 0.20 $\%$ of native Ti impurities \cite{Hildebrand2017, Hildebrand2016, Hildebrand2014}. Clean surfaces were obtained after cleaving in ultrahigh vacuum at room temperature (RT). Potassium (K) atoms were evaporated \textit{in situ} from a carefully outgassed SAES getter source onto the freshly cleaved TiSe$_2$ surfaces kept at $\sim$40 K to inhibit K intercalation \cite{Rossnagel2010, Caragiu2005}. During the K evaporation, the pressure was maintained below 5 $\times$ 10$^{-10}$ mbar. The ARPES measurements were carried out using a Scienta Omicron DA$30$ photoelectron analyzer with monochromatised He-I and II radiations ($h\nu$=$21.2$ eV, $40.8$ eV, Specs GmbH) and laser excitation source ($h\nu$=$6.3$ eV, APE GmbH). The total energy resolution was 5 meV and the base pressure during experiments was better than 1.5 $\times$ 10$^{-10}$ mbar. 

Figure \ref{fig1}a) shows the evolution of the K 3$p$ shallow core level upon sequential K depositions on the 1$T$-TiSe$_2$ surface. For K coverages lower than 0.10 monolayer (ML) [1 ML corresponds to one K atom per 1$T$-TiSe$_2$ surface unit cell, see Supplemental Material (SM) \footnote{See Supplemental Material at url for details on the K coverage characterization, the density functional theory calculations, the experimental charge transfer determination, the slabs potentials extraction, the comparison of He-I and laser-based ARPES intensity maps at high K coverage, and the used self-consistent Poisson-Schr{\"o}dinger parameters, which includes Refs.\cite{kresse1993,kresse1994,kresse1996a,kresse1996b,blochl1994,kresse1999a, Perdew1996, Herath2020, Jaouen2014,Wilson1978b,Yukawa2015}}], the photoemission spectrum mainly consists of one component at 18.4 eV binding energy (BE) associated with dispersed K atoms on the surface. With increased coverages, the dispersed phase quickly saturates and a second component corresponding to closely-packed K atoms continuously grows on the low BE side (17.9 eV BE) and starts to be dominant at 0.45 ML \cite{Caragiu2005}. The comparison of the 12$\times$12 nm$^2$ constant-current STM images of the pristine [Fig. \ref{fig1}b)] and of a 0.33 ML K-covered surface [Fig. \ref{fig1}c)] \footnote{Here, the K adlayer was obtained by a single evaporation on a freshly cleaved 1$T$-TiSe$_2$ surface kept at 40 K in the ARPES chamber before being quickly transferred in the STM previously cooled down to 4.6 K. We cannot thus exclude that the sample experienced a small temperature cycling during the transfer and that a tiny fraction of K atoms has intercalated. Also note that the CDW modulations gets smeared out in STM images taken at V$_{\text{bias}}=-1$ V because many electronic states that are not concerned with the electronic band structure reconstruction are taken into account.}, shows that the dispersed and close-packed forms of the K atoms are moreover randomly distributed, without surface reconstructions, as confirmed by the corresponding low-energy electron diffraction (LEED) measurements showing the typical 1 $\times$ 1 hexagonal pattern with an increased diffuse background for the K-covered surface. Figures \ref{fig1}d) and \ref{fig1}e) respectively show large-energy scale ARPES intensity maps taken at T = 40 K along the $\bar{M}$-$\bar{\mathrm{\Gamma}}$-$\bar{M}$ and $\bar{\mathrm{\Gamma}}$-$\bar{M}$-$\bar{\mathrm{\Gamma}}$ directions of the surface Brillouin zone (BZ) [Fig. \ref{fig1}f)] of the pristine (left-hand side), and the 0.14 ML K-covered (right-hand side) 1$T$-TiSe$_2$ surfaces. Density-functional theory (DFT) band structures calculations obtained from 6-layers slabs are presented for comparison [Fig. \ref{fig1}g)] (see SM).
A theoretical 0.125 ML K-covered 1$T$-TiSe$_2$ surface has been considered using a 2 $\times$ 4 K adlayer. Upon K adsorption, the two outer Se-Ti-Se layers on each side of the slab have been allowed to relax whereas the two inner ones have been fixed to the relaxed positions of the CDW phase of pristine 1$T$-TiSe$_2$. As expected from the highly electropositive character of K, the charge density difference indicates that upon adsorption the TiSe$_2$ surface layer becomes strongly electron-doped. According to our Bader analysis \cite{Tang2009}, each K adatom gives 0.7 $e^-$ to the first Se layer, in good agreement with our experimentally value of $\sim$0.6 $e^-$/K atom (see SM). Overall, the DFT-calculated unfolded band dispersions reproduce the main spectroscopic features of the ARPES intensity maps of both the pristine and the K-doped surfaces very well, namely, the bands binding energies, the effective masses, as well as the spectral weights [Fig. \ref{fig1}d)-e)]. In the pristine CDW phase, the Ti 3$d_{z^2}$ electron band of non-bonding character at $\bar{M}$ determines the low-energy physics of 1$T$-TiSe$_2$ \footnote{We refer to the crystallographic reference frame with $z$ out of the plane for the orbitals projection}, but is decoupled from the CDW energetics since it lies within the CDW gap which is opened at higher energies between the hybridized Se 4$p_{x,y}$ hole band and the so-called "Mexican–hat"-shaped unoccupied electron band [see the vertical red arrows on Fig. \ref{fig1}d)-e)] \cite{Watson2020, Watson2019,Jaouen2019}. Additionally, DFT tracks the main effects experimentally observed upon K adsorption. The Ti 3$d_{z^2}$ electron band at $\bar{M}$ is shifted below the Fermi level ($E_F$) and backfolded hole bands characteristic of the 2 $\times$ 2 CDW order are present. Looking more closely at the changes both in the experimental and DFT band structures, we see that the K-adsorption gives rise to new electronic states at $\bar{\mathrm{\Gamma}}$ and $\bar{M}$ [see the horizontal black arrows on Fig. \ref{fig1}d)-e)] that seem to have been pulled down from the Se 4$p_{x,y}$, Ti 3$d_{z^2}$, and Mexican-hat bulk-band edges, and whose origin will now be addressed. 

\begin{figure}[t]
\includegraphics[scale=1]{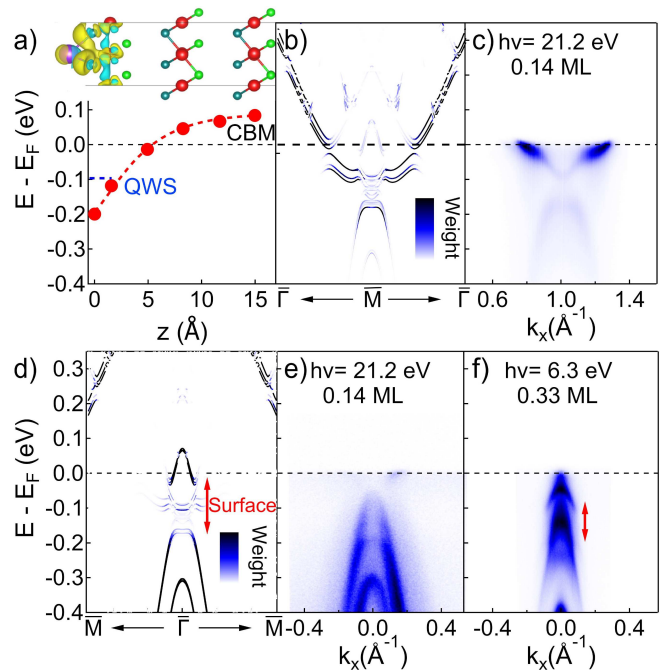}
\caption{a) Band-bending potential extracted from DFT (see text and SM). b) Near-$E_F$ zoom-in of the surface layer-projected DFT-calculated band structure along $\bar{\mathrm{\Gamma}}$-$\bar{M}$-$\bar{\mathrm{\Gamma}}$ of the 0.125 ML K-covered slab. The color scale from white to dark blue indicates the spectral weights obtained using the unfolding procedure (see SM). c) ARPES intensity map at $\bar{M}$ of a 0.14 ML K-covered surface. $h\nu$=$21.2$ eV, T = 40 K. d) Same as b) along $\bar{M}$-$\bar{\mathrm{\Gamma}}$-$\bar{M}$. The red arrows indicates the surface CDW gaps. e) Same as c) at $\bar{\mathrm{\Gamma}}$. f) Laser-ARPES intensity map taken at $\bar{\mathrm{\Gamma}}$ of a 0.33 ML K-covered surface. The red arrow indicates the surface CDW gap. $h\nu$=$6.3$ eV, T = 40 K.}\label{fig2}
\end{figure}

Firstly, our layer-projected DFT band structures reveal that they are localized in the topmost surface layer without any sign of interlayer hybridization (see SM), demonstrating their purely 2D character. Next, similarly to a DFT study of K-covered Bi$_2$Se$_3$ surfaces \cite{Forster2015}, we have extracted the change induced by the K adsorption in the laterally averaged effective potential of the slab (see SM). The resulting potential exhibits two contributions, an adsorbate-specific surface dipole layer responsible for the strong work function decrease of $\sim$3.6 eV at the highest coverage (see SM), superimposed on a more long-range band-bending potential of $\sim$0.3 eV in depth that extends over roughly 15 \AA~into the bulk [Fig. \ref{fig2}a)], and acts as a near-surface confinement well for electrons. As we confirm later, this leads to the formation of a 2DEG at the surface and to the emergence of quantum-well sub-band states (QWS) such as the one of the original $k_z$-dispersing bulk Ti 3$d_{z^2}$ state at $\sim$0.1 eV BE [Fig. \ref{fig2}b)], that appears sharp and with a strong spectral weight in experiment [Fig. \ref{fig2}c)]. In the same way, the Mexican-hat shaped band that crosses $E_F$ and the new hole band which lies at $\sim$0.15 eV BE at $\bar{\mathrm{\Gamma}}$ [Fig. \ref{fig2}d)], originate from surface quantum confinement of the bulk CDW states. Consistent with this, the latter is backfolded to $\bar{M}$ [Fig. \ref{fig2}b)], and has its experimental counterparts in the ARPES intensity maps at $\sim$0.20 eV BE both at $\bar{M}$ and $\bar{\mathrm{\Gamma}}$ [Fig. \ref{fig2}c)-e)].       

Importantly, compared to the Ti 3$d_{z^2}$-derived QWS whose energy shift with respect to its associated bulk-band only depends on the 2DEG density, those of the CDW-related QWS also include the effects of a reduced amplitude of the surface periodic lattice distortion (PLD) and the associated Ti$-d$ and Se$-p$ hybridization. This is manifested in our DFT calculations by the smaller band gap of the surface-confined CDW band gap compared to the bulk [see the red arrow on Fig. \ref{fig2}d)]. In the experiment, we cannot access the CDW gap  since the Mexican-hat mainly falls in the unoccupied part of the band structure. Probing a more highly electron-doped surface covered by 0.33 ML of K by laser-based ARPES which, compared to He-I, is not hindered by the increased inelastic background of electrons [Fig. \ref{fig2}f)] (see SM), allows us to reveal the full low-energy electronic structure of the confined CDW phase with a strongly reduced band gap completely shifted below $E_F$ [red arrow on Fig. \ref{fig2}f)]. 

\begin{figure}[b]
\includegraphics[scale=1]{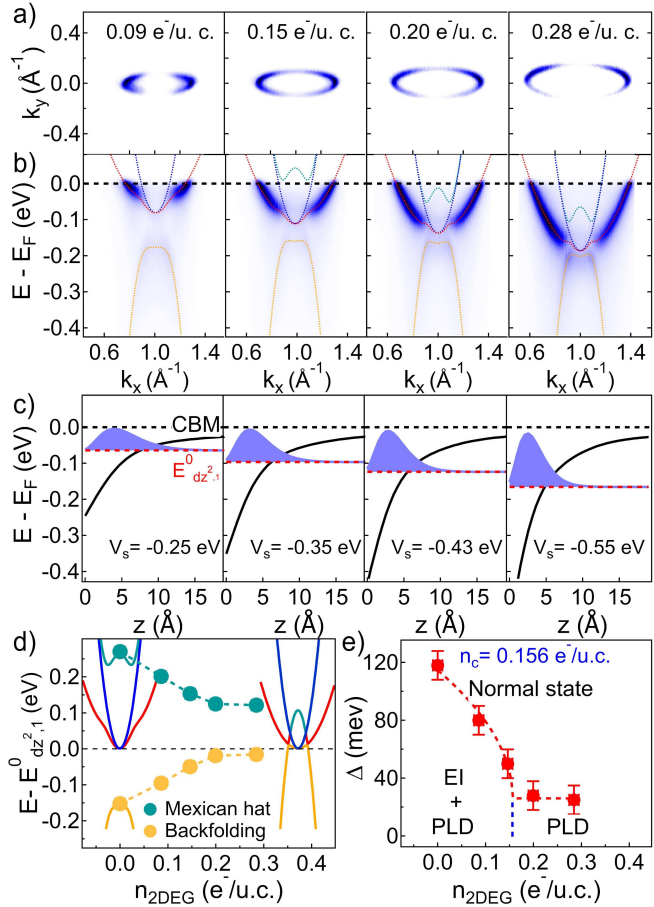}
\caption{a) FS at $\bar{M}$ for increasing $n_{2DEG}$. $h\nu$=$21.2$ eV, T = 40 K. b) Corresponding ARPES intensity maps. The near-$E_F$ dispersions calculated within the effective Hamiltonian model of four interacting bands are also superimposed. c) PS-calculated near-surface band bendings (black lines), eigenenergies $E^{0}_{dz^{2},1}$ (red-dashed lines) and modulus-squared of the eigenfunctions $\vert\phi^{0}_{dz^{2},1}(z)\vert^2$ (blue areas) of the first QWS as a function of $n_{2DEG}$. The surface potential values, $V_S$, are also indicated. d) Evolution of the BEs of the Mexican hat (green bands) and backfolding (orange bands) taken with respect to  $E^{0}_{dz^{2},1}$ as a function of $n_{2DEG}$. Sketches of the bands configuration in the unperturbed CDW phase and surface-confined semimetallic normal state are also shown. e) $\Delta$ as a function of $n_{2DEG}$. The red-dashed line is a mean-field like fitting with $\Delta (n_{2DEG}) = \Delta_{0}(1-n_{2DEG}/n_c)^{1/2} +\Delta_{off.}$ with $\Delta_{0}=$88$\pm$24 meV, $\Delta_{off.}=$26$\pm$15 meV and $n_c=$0.156 $\pm$0.024 $e^{-}/u.c.$.}\label{fig3}
\end{figure}

At this stage, we have shown that the K-covered 1$T$-TiSe$_2$ surface can host a 2DEG coexisting with a surface-confined CDW and the experimental and theoretical tendencies of electron doping to destabilize the CDW. Nevertheless, at the level of our considered exchange-correlation functional, the DFT calculations are unable to generate, already for the pristine CDW phase, the electronic bands, the atomic structure and PLD amplitude in simultaneous agreement with experiments \cite{Bianco2015}. It has been shown that only an inclusion of the exact exchange interaction within hybrid functionals properly captures the full properties of TiSe$_2$ \cite{Hellgren2017}. Therefore any attempt to drive some conclusions about the contribution of the excitonic and structural facets of the CDW state from our DFT calculations is irrelevant. Instead the unprecedented opportunity to tune the free-carrier density within the 2DEG opens an appealing way for probing the perturbations of the electronic and lattice subsystems in reduced dimensionality upon Coulomb screening. 

Figures \ref{fig3}a) and \ref{fig3}b) show the evolution of the 1$T$-TiSe$_2$ Fermi surfaces (FS) and the corresponding ARPES intensity maps at $\bar{M}$ upon increased K coverage obtained by cumulative evaporations (from left to right). The electron density in the 2DEG, $n_{2DEG}$, has been obtained from the Luttinger area of the FS, Fig \ref{fig3}a), and found to linearly increase with the K coverage (see SM for details). The order parameter, $\Delta$, of the surface-confined CDW has been determined by adjusting near-$E_F$ dispersions calculated within an effective Hamiltonian model of four interacting bands \cite{Jaouen2019, Monney2009a} to those of the experimental Ti 3$d_{z^2}$ and CDW-hybridized QWS, Fig \ref{fig3}b) (see SM) \footnote{Within our 4-band model we thus do not consider the two spin-orbit split Se 4$p$ hole bands that lead to a one non-bonding state and a pair of non-degenerated bonding states upon orbitals hybridization allowed by the PLD.  In our case, we obtained one doubly-degenerate bonding state (the Mexican hat) and one non-bonding Ti 3$d_{z^2}$ state (red-dashed line), the spectral weight of the blue-dashed dispersion being exactly zero for every $k$ along the $\bar{\mathrm{\Gamma}}$-$\bar{M}$ direction \cite{Monney2009a}. The two hole-bands of TiSe$_2$ being well separated in energy with only one cutting $E_F$ in the normal state, we expect the effect of the second hole band to be a minor correction to our obtained order parameter values \cite{Chen2018}.}. In contrast to the CDW-related states, we first see that the Ti 3$d_{z^2}$-derived QWS monotonoulsy shifts towards higher BE with $n_{2DEG}$ [dashed-red bands on Fig. \ref{fig3}b)]. As previously mentioned, this is because its BE only depends on the depth of the surface quantum well and is completely unrelated to the CDW energetics. This is confirmed by our self-consistent PS calculations that intentionally omit the CDW-hybridized states and used $n_{2DEG}$ as the only varying parameter (see SM for details on the PS calculations). As seen Fig. \ref{fig3}c), the obtained eigenenergies, $E^{0}_{dz^{2},1}$, (red-dashed lines), of the first bound QWS (blue areas that show the modulus-squared of the corresponding eigenfunctions $\vert\phi^{0}_{dz^{2},1}(z)\vert^2$) almost perfectly match the experimental ones for all considered $n_{2DEG}$ and related surface band-bending potentials $V_S$. Let us now discuss the impact of the increasing $n_{2DEG}$ on the CDW-hybridized QWS. Figure \ref{fig3}d) shows the evolution of the BEs of the top of the Mexican-hat [dashed-green bands in Fig. \ref{fig3}b)] and backfolded hole bands [dashed-orange bands in Fig. \ref{fig3}b)] obtained from the model with respect to  $E^{0}_{dz^{2},1}$ as a function of $n_{2DEG}$. With increased $n_{2DEG}$, they symmetrically approach each other, manifesting the closure of the confined CDW gap towards a surface-confined semimetallic normal state.

Interestingly, $\Delta$ exhibits a rapid drop as a function of $n_{2DEG}$ [Fig. \ref{fig3}e)], that follows a mean-field like behavior up to a critical electron density $n_c=$0.156 $\pm$0.024 $e^{-}/u.c.$. Within the EI scenario of the phase transition when the shift of the chemical potential, initially placed at mid-gap between the hole an electron bands, exceeds half of the excitonic binding energy described by a Bardeen-Cooper-Schrieffer-like order parameter, the electron band starts to be filled and the EI phase becomes unstable \cite{Chen2020}. In 1$T$-TiSe$_2$, the CDW gap opens on both sides of the Ti 3$d$-Se 4$p_{x,y}$ electron-hole band crossings and enlarges around the new Ti 3$d_{z^2}$ non-bonding state which is unaffected by the orbital-selective hybridization allowed by the PLD \cite{Watson2020}. Therefore, our experimental finding of a critical density at which the shift of the Ti 3$d_{z^2}$-derived QWS exceeds half of $\Delta (0)$ and the upper edge of the confined CDW gap, i.e., the mexican-hat, goes below $E_F$, strongly suggest that the gap shrinking of the surface-confined CDW relates to a decrease of an excitonic binding energy. 

Note, however, that for the highest $n_{2DEG}$, $\Delta$ has a non-zero value. Indeed, a small hybridization gap is clearly opened below $E_F$ at $\bar{M}$ and the backfolded hole band still appears sharp [see the right-hand side panel of Fig. \ref{fig3}b)]. This indicates that both the CDW amplitude and phase fluctuating modes at the surface are gapped \cite{Chen2018}, i.e. the persisting surface-confined CDW is commensurate with long-range phase coherence.
Hence, we argue that the small remaining CDW gap at dopings where excitonic interactions cannot be at play originates solely from a persisting 2 $\times$ 2 structural order  that could be maintained by 
electrostatic interaction with the bulk which hosts a perfectly locked-in CDW. 

At a first sight, our critical electron density $n_c$ seems to overshoot the commonly accepted one by a factor of $\sim$3 \cite{Watson2020, Rossnagel2010, Zhao2007, Morosan2006a}. Nevertheless, several other dispersed $n_c$ values have been reported in the literature (see for example Table I in Ref.\cite{Rossnagel2010}), depending essentially of the means of doping and the use of local or spatially-averaging techniques for probing the CDW \cite{Novello2017, Yan2017, Kogar2017, Spera2019}, that shows a natural tendency towards nanoscale phase separation in crystals doped in the high-temperature phase \cite{Jaouen2019}. In our case, the ordered phase of TiSe$_2$ is electron-doped by filling the electron band of non-bonding character that further allows to solely probe the impact of increased Coulomb screening. This not only does not give rise to a significant energetic penalty compared with the equivalent filling of the electron band in the undistorted phase \cite{Watson2020}, but also allows to maintain the CDW long-range phase coherence. It turns out that our reported critical doping could actually appear as the most representative one to date. It is also interesting to note that it is very close to that associated with the emergence of superconductivity in electrically-gated 1$T$-TiSe$_2$ nanosheets \cite{Li2015}. This suggests that the superconducting phase could compete with the excitonic contribution of the CDW, thus pointing to an unconventional pairing mechanism. Yet, in our case, CDW domain walls that have been proposed to host superconductivity are missing \cite{Yan2017, Kogar2017, Li2015}. Therefore, the questions about the nature of the superconducting phase in bulk doped 1$T$-TiSe$_2$, its existence at the 2D limit, and of their possible unconventional origin remain open. 

To conclude, through the specific case of 1$T$-TiSe$_2$, our study reports coexisting 2DEG and quantum-confined CDW at a TMD surface and constitutes a prime example of a carrier-density controlled many-body quantum state in reduced dimensionality. This approach has allowed us to completely nullify an excitonic energy gain at the surface and at thermal equilibrium while maintaining a 2D CDW order with long-range phase coherence that we thereby propose to be purely electron-phonon coupled. We believe that our study further provides experimental avenues for controlling dimensional crossovers within unexplored parts of the electronic phase diagrams of correlated materials.

\begin{acknowledgments}
This project was supported by the Fonds National Suisse pour la Recherche Scientifique through Div. II. Skillful technical assistance was provided by F. Bourqui, B. Hediger and O. Raetzo.
\end{acknowledgments}

\bibliographystyle{apsrev4-2}
\bibliography{library1}

\end{document}